\documentclass[12pt]{article}
\usepackage[dvips]{graphicx, color}
\usepackage{amsfonts}
\usepackage{amsmath}
\usepackage{amscd}
\usepackage{amssymb}
\usepackage{bm}
\usepackage[all]{xy}
\usepackage{enumerate}
\newcommand{\GSM}{G_{\textrm{SM}}}
\newcommand{\GpSM}{G'_{\textrm{SM}}}
\newcommand{\GbSM}{\overline{G}_{\textrm{SM}}}

\def\PRD#1#2#3{Phys.\ Rev.\ \textbf{D {#1}},  #2 (#3)}
\def\PRL#1#2#3{Phys.\ Rev.\ Lett.\ \textbf{{ #1}},  #2 (#3)}

\begin{document} 

\begin{flushright}

\end{flushright}

\vspace{3mm}

\begin{center}
{\large \bf  Quantization of hypercharge  in  gauge groups\\
locally isomorphic  but globally nonisomorphic 
to $SU(3)_{\rm c} \times SU(2)_{\rm L}\times U(1)_{\rm Y}$ }

\vspace{15mm}

Takaaki Hashimoto, 
 
Mamoru Matsunaga, 
             \footnote{corresponding author}
             \footnote{E-mail: matsuna@phen.mie-u.ac.jp} and
Kenta Yamamoto 
\end{center}

\begin{center}
\textit{
Department of Physics Engineering, Mie University, \\
     Tsu 514-8507, JAPAN
 }
\end{center}

\vspace{5mm}

\begin{abstract}
In the Standard Model the hypercharges of quarks and leptons are not determined  by the gauge group itself. 
In a recent paper  [C.~Hattori \textit{et al}. \PRD{83}{015009}{2011}] it is shown that, if the direct product gauge group $\GSM$ is slightly modified to the semidirect product group $\GpSM$,  hypercharges are restricted to quantized values as  $n/6 \mod {\mathbb{Z}}\;(n = 0,1,3,4) $. In this brief paper, we examine  all of the compact Lie groups locally isomorphic to $\GSM$, and show that $\GpSM$ (or its isomorphisms)  is the unique possibility that yields the correct hypercharge quantization.
\end{abstract}

\newpage 
The  Standard Model (SM) of elementary particles is based on the direct product group  
$\GSM = SU(3)_{\rm c} \times SU(2)_{\rm L} \times U(1)_{\rm Y}$ and adopts the assignment of representations of $\GSM$ to fermions as
 $q_\textrm{L} = (\mathbf{3}, \ \mathbf{2}, \ 1/6)$, $u_\textrm{R} =(\mathbf{3}, \ \mathbf{1}, \ 2/3)$, $d_\textrm{R} = (\mathbf{3}, \ \mathbf{1}, \ -1/3)$, 
$l_\textrm{L}  = (\mathbf{1}, \ \mathbf{2}, \ -1/2)$, and $e_\textrm{R} = (\mathbf{1}, \ \mathbf{1}, \ -1)$. 
These peculiar values of the hypercharges $y$  ($U(1)_{\rm Y}$-charges) in this assignment  are fixed by the phenomenological requirement of the extremely precise equality of the absolute values of electric charges of protons and electrons. 
These values are not determined by the gauge group $\GSM $ itself.
It is well known that they are fixed by an extension of the gauge group to $SU(5)$, and this fact is one of the motivations to consider this larger group \cite{SU(5)}.

Recently, in Ref.~\cite{HMM}, it was shown that, by choosing the semidirect product $\GpSM = [SU(3)_{\rm c} \times SU(2)_{\rm L}] \rtimes U(1)_{\rm Y}$ without a group extension, we can restrict hypercharges to quantized values as  $n/6 \mod {\mathbb{Z}}\;(n = 0,1,3,4) $ within its linear representations. 
The requirement of gauge-anomaly cancellation further constrains the values of $y$'s. (For related works on the anomaly cancellation and the values of $y$, see  Refs.~\cite{ Weinberg, Marshak,Minahan}.)
It was also shown that  $\GpSM$ is isomorphic to some subgroup $G^{(5)}$ of $SU(5)$ and to the factor group $\GSM/\mathbb{Z}_6$, the cyclic group $\mathbb{Z}_6$ being generated by $(\omega_3 \mathbf{1}_3, \omega_2 \mathbf{1}_2, \omega_6) \in \GSM$, where $\omega_n$ is the primitive $n$th root of unity.
They are locally isomorphic but globally nonisomorphic to $\GSM$.

In this brief paper, we examine all of the compact Lie groups locally isomorphic to $\GSM$,  and show the semidirect product group $\GpSM$ (or its isomorphisms)  is the unique possibility that yields correct hypercharge quantization.

First, we recall a basic theorem of Lie group theory \cite{Pontryagin}, which states that, in general, any connected Lie group $G$ locally isomorphic to some group $G_0$ is a factor group of its  universal covering group $ \overline{G}_0$ by a discrete subgroup $\varGamma$ of the center $C( \overline{G}_0)$,  $G = \overline{G}_0/\varGamma$, and furthermore the fundamental group of $G$ is isomorphic to $\varGamma$.  As to $\GSM$, the universal covering group  is 
\begin{equation}
\GbSM= SU(3)_{\rm c} \times SU(2)_{\rm L} \times \mathbb{R},
\end{equation}
 $\mathbb{R}$ being the additive group of real numbers, and its center is $C (\GbSM)= \langle \omega_3 \mathbf{1}_3 \rangle \times \langle \omega_2 \mathbf{1}_2 \rangle \times \mathbb{R}$,
where  $\langle a \rangle$ denotes the cyclic group generated by the element $a$ of $C (\GbSM)$.

For the factor group
\begin{equation}
G = \GbSM/ \varGamma
\end{equation}
to be compact, $\varGamma$ must be an infinite group. Noting that a discrete subgroup of $\mathbb{R}$ can be written, with some real numbers $\alpha_i \neq 0$, as $\varGamma_\mathbb{R} = \sum_i n_i \alpha_i\; (n_i \in \mathbb{Z})$, we find that $\varGamma$ is $\langle \omega_3 \mathbf{1}_3 \rangle \times \langle \omega_2 \mathbf{1}_2 \rangle \times \varGamma_\mathbb{R} $ or its subgroup of infinite order.
If $\varGamma_{\mathbb{R}}$  contains several generators $\alpha_i$, they must be written as $\alpha_i = \alpha r_i$ with some real number $\alpha$ and rational numbers $r_i$. Otherwise the points identified by modding $\varGamma$ would distribute densely in $\mathbb{R}$ and the factor group $\GbSM/\varGamma$ would not form a Lie group. 
For the same reason, the number of  independent generators $\alpha_i $ must be finite.
Furthermore elementary arithmetic shows that the set $\{\sum_{i=1}^{I}  n_i r_i \bigm|n_i\in \mathbb{Z} \}$ can be rewritten, with some rational number $r$,   as $\{nr \,| \, n\in \mathbb{Z} \}$. 
Using the group isomorphism $\mathbb{R} \ni \theta \mapsto  (\alpha r/2\pi)\theta \in \mathbb{R}$, we can finally write, without loss of generality, as
\begin{equation}
\varGamma_\mathbb{R} = \{{ 2\pi n \,| \, n\in \mathbb{Z} }\}.
\end{equation}
In order to list  all possible $\varGamma$, we  invoke the fundamental theorem about a finitely generated Abelian group \cite{Rotman}. It means in our case that $\varGamma$ can be written as a product of finite or infinite order cyclic subgroups $\varGamma_i$'s as 
\begin{equation}
\varGamma = \varGamma_1  \varGamma_2 \cdots \varGamma_J,  \label{GGG}
\end{equation}
where  $\varGamma_i \bigcap  \varGamma_j =\{e \} \; (i \neq j)$ and $\varGamma_i   \varGamma_j $ denotes the set of all the elements $c_i c_j$  with  $c_i \in \varGamma_i$ and $c_j \in \varGamma_j$.

A candidate of generator $a_i$ of $\varGamma_i $ takes the form $( {\omega_3}^l \mathbf{1}_3 ,  {\omega_2}^m \mathbf{1}_2;\,n)$ with $l= 0, 1, 2; \, m = 0, 1;\, n \in \mathbb{Z}$.
Examples of $a_i $ are enumerated below:
\begin{alignat*}{4}
a_1 &:=  ( \omega_3 \mathbf{1}_3,  \mathbf{1}_2; 1 )\quad\quad
&\varGamma_1 &= \langle a_1\rangle = \{ (({\omega_3}^n \mathbf{1}_3,\, \mathbf{1}_2; \,n)\; | \; n \in \mathbb{Z} \} \\
a_2 &:=  ( \mathbf{1}_3, \omega_2 \mathbf{1}_2; 1)\quad\quad
&\varGamma_2 &= \langle a_2\rangle = \{ (\mathbf{1}_3,\,  {\omega_2}^n \mathbf{1}_2; \,n)\; | \; n \in \mathbb{Z} \} \\
a_3 &:=  ( \mathbf{1}_3,  \mathbf{1}_2; 1)\quad\quad   
&\varGamma_3 &= \langle a_3\rangle = \{ (\mathbf{1}_3,\, \mathbf{1}_2; \,n)\; | \; n \in \mathbb{Z} \} \\
a_4 &:=  (\omega_3 \mathbf{1}_3, \omega_2 \mathbf{1}_2; 1)\quad\quad   
&\varGamma_4 &= \langle a_4\rangle = \{ ({\omega_3}^n \mathbf{1}_3,\,  {\omega_2}^n \mathbf{1}_2; \,n)\; | \; n \in \mathbb{Z} \} \\
a_5 &:=  ({\omega_3}^2 \mathbf{1}_3, \omega_2 \mathbf{1}_2; 1)\quad\quad   
&\varGamma_5 &= \langle a_5\rangle = \{ ({\omega_3}^{2n} \mathbf{1}_3,\,  {\omega_2}^n \mathbf{1}_2; \,n)\; | \; n \in \mathbb{Z} \}. 
\end{alignat*}
The product $\varGamma_1 \varGamma_2$ should not be confused  with $\varGamma_4$.

We now consider linear representations $\mathcal R$ of $G$.
For any representation $\mathcal R$, the composition $\mathcal R\circ \pi$ forms a representation of  $\GbSM$. Conversely, each representation $\overline{\mathcal R}$ of $\GbSM$ that satisfies, for any $a \in \varGamma$, 
\begin{equation}
\overline{\mathcal R} (ga) =\overline{\mathcal R}(g) \quad (g \in \GbSM)   \label{inv_repr_1},
\end{equation}
induces a representation of $G$.
\centerline{
	\xymatrix{
	\GbSM\ar[dr]^{\overline{\mathcal R}} \ar[d]_{\pi} \\
	G \ar[r]_{\mathcal R\quad} & GL(V)  }
}

A representation of $\GbSM\ni \bigl(g_3, g_2, \theta\bigr)$ takes the tensor product form 
\begin{equation}
   \overline{\mathcal{R}}\bigl(g_3, g_2, \theta\bigr)
       = R _3\left(g _3\right) 
         \otimes R _2\left(g _2\right) 
            \otimes e^{i y \theta},
	\label{rep32}
\end{equation}
where $R_3 \,\left( R_2\right)$ is an irreducible representation of $SU(3)_{\rm c} \,\left( SU(2)_{\rm L} \right)$. They are specified by the Dynkin label pair $[ \lambda_1, \lambda_2 ]_D$ (isospin $I$).

Thus, in order to find  all the possible representations of $G = \GbSM/ \varGamma$, we only have to choose, among  $ \overline{\mathcal{R}}$ given in Eq.~\eqref{rep32}, representations that satisfy Eq.~\eqref{inv_repr_1}.
Since $\varGamma$ is a product of the cyclic groups  $\varGamma_i = \langle a_i \rangle$ as shown in Eq.~\eqref{GGG}, the invariance condition Eq.\eqref{inv_repr_1} is equivalent to  
\begin{equation}
\overline{\mathcal R} (a_i ) = \mathbf{ 1}   \label{inv_repr_2} ,
\end{equation}
i.e.  each generator $a_i$ gives a selection rule of the representations of $\GbSM$.

For example, if we take $a_1 =  (\omega_3\mathbf{1}_3,  \mathbf{1}_2; 1)$, Eq.~\eqref{inv_repr_2} becomes
\begin{equation*}
    \overline{\mathcal{R}}\bigl(a_1)
       = e^{2\pi i\frac{r_3}{3}} 
                           e^{2\pi iy}\mathbf{ 1} = \mathbf{ 1} .
	\label{rep321}
\end{equation*}
This implies the selection rule $y \equiv  - {r_3}/{3} := - (\lambda_1 + 2 \lambda_2)/3\mod {\mathbb{Z}}$, which contradicts the SM assignment of hypercharge.
In a similar way, $a_2 =  ( \mathbf{1}_3,  \omega_2\mathbf{1}_2; 1)$ as well as $a_3 = ( \mathbf{1}_3,  \mathbf{1}_2; 1)$ give  incorrect selection rules  $y \equiv  {r_2}/{2} := I $ and $ y \equiv 0\mod {\mathbb{Z}}$, respectively.
 On the other hand the choice  $a_4 =   (\omega_3 \mathbf{1}_3, \omega_2 \mathbf{1}_2; 1)$  gives  
\begin{equation}
	y \equiv \frac{r_2}{2} - \frac{r_3}{3} \mod {\mathbb{Z}} \label{Qy},
\end{equation}
which is consistent with the SM assignment, while $a_5 =  ({\omega_3}^2 \mathbf{1}_3, \omega_2 \mathbf{1}_2; 1)$ gives the wrong sign hyperchages. (We note in passing the complex conjugation of $SU(3)$, $g_3 \mapsto {g_3}^\ast$, gives the isomorphism  $\GbSM/\varGamma_5\cong \GbSM/\varGamma_4$.)

In this way we can see that all the choices of generators other than $a_4$ give incorrect hypercharges. 
(The reader might think that, for example,  $\langle {a_3}^6 \rangle$  gives the selection rule consistent with the SM. Notice, however, $\langle {a_3}^6 \rangle$ is a subgroup of $\langle {a_4} \rangle$ because ${a_3}^6 = {a_4}^6$.)
Thus we find the group $\varGamma$, which gives the correct selection rule for hyperchrges, to be 
\begin{equation}
\varGamma_4 = \langle  (\omega_3 \mathbf{1}_3, \omega_2 \mathbf{1}_2; 1)\rangle.
\end{equation}

The factor group
\begin{equation}
\GbSM/\varGamma_4 =[SU(3)_{\rm c} \times SU(2)_{\rm L} \times \mathbb{R}]/ \langle  (\omega_3 \mathbf{1}_3, \omega_2 \mathbf{1}_2; 1)\rangle
\end{equation}
is seemingly different from the factor group
\begin{equation}
\GSM/\mathbb{Z}_6 =  [SU(3)_{\rm c} \times SU(2)_{\rm L} \times U(1)]/\langle  (\omega_3 \mathbf{1}_3, \omega_2 \mathbf{1}_2, \omega_6)\rangle,
\end{equation}
the latter of which is shown to be isomorphic to $\GpSM$ in Ref.~\cite{HMM}. They are, in fact, isomorphic to each other.
Indeed, by rewriting as
\begin{alignat*}{3}
\varGamma_4 &= \langle  (\omega_3 \mathbf{1}_3, \omega_2 \mathbf{1}_2; 1)\rangle \\
 &=\{ \left({\omega_3}^{n + 6m} \mathbf{1}_3, {\omega_2}^{n + 6m} \mathbf{1}_2; n + 6m\right)\; | \; m, n \in \mathbb{Z} \} \\
 &=\left\{ ({\omega_3}^{n} \mathbf{1}_3, {\omega_2}^{n} \mathbf{1}_2; \, 6(\frac{n}{6} + m))\; | \; m, n \in \mathbb{Z} \right\} 
 \end{alignat*}
 and by considering the map $\theta \mapsto 6\theta$, we obtain, with $\varGamma^{\prime}_{4} := \{ ({\omega_3}^{n} \mathbf{1}_3, {\omega_2}^{n} \mathbf{1}_2; \, n/6 \,+ m)\; | \; m, n \in \mathbb{Z} \} $,
\begin{equation*}
\GbSM/\varGamma_{4} \cong \GbSM/\varGamma^{\prime}_{4} ,
\end{equation*}
whose right-hand side is nothing but $\GSM/\mathbb{Z}_6$.
\\

We conclude that the semidirect product group $\GpSM$ (or its isomorphisms)  is the unique possibility that yields correct hypercharge quantization.
This result  gives no practical significance to the SM physics, except for predictions about exotic matter fields, which have already been given in Ref.~\cite{HMM}. However, the reasoning given here will serve the study of beyond-SM gauge groups and matter contents.
\\
\\
One of the authors (M.M.) wishes to express his gratitude to  Chuichiro Hattori and Takeo Matsuoka for several helpful comments concerning the structure of $\varGamma_\mathbb{R}$ and the final result.



\end{document}